# Finite series representation for the bound states of a spiked isotropic oscillator with inverse-quartic singularity

A. D. Alhaidari

*Saudi Center for Theoretical Physics, P.O. Box 32741, Jeddah 21438, Saudi Arabia*

**Abstract:** We use the tridiagonal representation approach to obtain an exact solution of the three-dimensional radial Schrödinger equation for a spiked oscillator with inverse quartic singularity and for all angular momenta. The solution is a finite series of square integrable functions written in terms of the Bessel polynomial.

## 1. Introduction

In the physics literature, most studies of the isotonic oscillator are concerned with special cases or variants of the following five-parameter potential

$$V_{\mathrm{I}}(x)=\frac{1}{2}\omega^2x^2+\frac{gx^2+bx\pm c^2}{\left(x^2+a^2\right)^2}, \tag{1}$$

where $\{a,b,c\}$ are length parameters, $g$ is a dimensionless coupling and $\omega$ is the oscillator frequency. This is a confining potential, which is regular (locally non-singular) unless $a=0$. Two versions of this oscillator in one dimension that were studied extensively have the following forms (see, for example, [1-3] and references therein):

$$V_{\mathrm{II}}(x)=V_0\left(\frac{x}{x_0}-\frac{x_0}{x}\right)^2, \tag{2a}$$

$$V_{\mathrm{III}}(x)=\frac{1}{2}\omega^2x^2+g\,\frac{x^2-a^2}{\left(x^2+a^2\right)^2}. \tag{2b}$$

$V_{\mathrm{II}}(x)$ is inverse-square singular at the origin. However, $V_{\mathrm{III}}(x)$ is regular unless $a=0$. In three dimensions with spherical symmetry, the following isotonic oscillator potential (including the kinetic energy orbital term) was also considered:

$$V_{\mathrm{IV}}(r)=\frac{\ell(\ell+1)}{2r^2}+\frac{1}{2}\omega^2r^2+g\,\frac{r^2-a^2}{\left(r^2+a^2\right)^2}. \tag{3}$$

where $\ell$ is the angular momentum quantum number (see, for example, [4] and references therein). Interest in these isotonic oscillator potentials and their applications in physics are outlined in the cited literature above. The generalized potential model (1) has interesting applications in atomic, molecular and nuclear physics. With $a \neq 0$, it extends the simple harmonic oscillator potential model to fit most of the physically relevant modifications to the oscillatory motion with a four-parameter fitting control. However, the most dramatic extension comes into play if $a = 0$ where then the potential becomes singular at the origin with a three-parameter control of the singularity strength. For a one-parameter singularity, sometimes such isotonic oscillator is referred to as the "spiked harmonic oscillator" [5-16]. For high orbital motion (large $\ell$ values), the effect of the singularity is diminished. However, for low orbital motion (small values of $\ell$), the effect of the singularity becomes more dramatic.

In this work, we consider the most singular version of the isotonic oscillator potential (1) (including the orbital term) that reads

$$V(r) = \frac{\ell(\ell+1)}{2r^2} + \frac{1}{2}\omega^2 r^2 + \frac{1}{2}\frac{a^2}{r^4}, \tag{4}$$

It is a special case of the potential (1) with $a \mapsto 0$, $b = 0$, $c \mapsto \frac{1}{\sqrt{2}}a$, and $g = \frac{1}{2}\ell(\ell+1)$. The exact solution (wavefunction and energy spectrum) of the Schrödinger equation with this potential is a highly non-trivial task. Consequently, this problem was addressed by many authors where very good approximations are obtained for the lower part of the energy spectrum using various techniques [5-16]. These studies produced excellent approximations of the solution in a given region of the potential parameter (and angular momentum) space. Several conventional methods were used in these studies in attempting to obtain exact solutions of the wave equation but with limited success. These methods include factorization, supersymmetry, shape invariance, point canonical transformation, Darboux transformation, asymptotic iteration, and the Nikiforov-Uvarov method. Nonetheless, we were able to obtain an exact solution using the novel "tridiagonal representation approach" (TRA) [17]. In sections 2 and 3, we formulate the problem and solve it within the TRA then present our results in sections 4 and 5. We tail the paper with a conclusion in section 6.

## 2. TRA formulation of the problem

In the atomic units $\hbar = m = 1$, the time-independent radial Schrödinger equation for the spiked oscillator potential (4) reads as follows

$$\left[-\frac{1}{2}\frac{d^2}{dr^2} + \frac{\ell(\ell+1)}{2r^2} + \frac{1}{2}\omega^2 r^2 + \frac{1}{2}\frac{a^2}{r^4} - E\right]\psi(r) = 0, \tag{5}$$

where $E$ is the energy and $\psi(r)$ is the corresponding radial wavefunction. Now, we employ the tools of the TRA in the formulation and solution of this problem. For detailed description of the TRA, one may consult [17,18] and references therein. In the TRA, we start by expanding the wave function in a series as $\psi(r) = \sum_n f_n(E)\phi_n(x)$, where $\{\phi_n(x)\}$ is a set of square integrable functions that produce a tridiagonal matrix representation for the wave operator (5). We choose the dimensionless variable $x$ as $x = 1/\omega r^2$. In terms of this variable, the wave equation (5) becomes

$$\mathcal{D}\psi(r) = -2\omega x\left[x^2\frac{d^2}{dx^2}+\frac{3}{2}x\frac{d}{dx}-\frac{1}{4}\ell(\ell+1)-\frac{1/4}{x^2}-\frac{1}{4}(\omega a^2)x+\frac{\varepsilon/2}{x}\right]\psi(r)=0. \quad (6)$$

where $\varepsilon = E/\omega$. A square-integrable basis set that could support a tridiagonal matrix representation for the wave operator $\mathcal{D}$ has the following elements

$$\phi_n(x) = x^{\alpha}e^{-1/2x}Y_n^{\mu}(x), \quad (7)$$

where $Y_n^{\mu}(x)$ is the Bessel polynomial on the positive real line whose relevant properties are given in the Appendix. The degree of the polynomial is limited by the negative parameter $\mu$ where $n = 0,1,2,..,N$ with *N* being the largest integer less than $-\mu-\frac{1}{2}$. Choosing $\alpha = \mu+\frac{1}{4}$ and using the differential equation of the Bessel polynomial (A4), we can evaluate the action of the wave operator on the basis elements giving

$$\mathcal{D}\phi_n(x) = -\frac{\omega}{2}x^{\mu+\frac{5}{4}}e^{-1/2x}\left[(2n+2\mu+1)^2-\left(\ell+\tfrac{1}{2}\right)^2-(\omega a^2)x+\frac{4\mu+2\varepsilon}{x}\right]Y_n^{\mu}(x). \quad (8)$$

The orthogonality of $Y_n^{\mu}(x)$ given by Eq. (A3) shows that a tridiagonal matrix representation $\langle\phi_m|\mathcal{D}|\phi_n\rangle$ is obtained if and only if $[...]Y_n^{\mu}(x)$ in (8) becomes a sum of terms proportional to $Y_n^{\mu}(x)$ and $Y_{n\pm1}^{\mu}(x)$ with constant factors. Hence, the three-term recursion relation (A2) dictates that the terms inside the square brackets is permitted to be only linear in *x*. Thus, the last term which proportional to $x^{-1}$ must vanish and thus we should choose the basis parameters $\mu$ as follows

$$2\mu = -\varepsilon = -E/\omega. \quad (9)$$

Since the basis parameter $\mu$ is required to be negative then the energy spectrum of the oscillator potential must be positive. In fact, it should be greater than $\omega$ since $\mu < -\frac{1}{2}$. This is consistent with the total confinement and positivity of the potential (4). Moreover, the number of bound states that could be obtained by our solution is unlimited since $\mu$ can assume any negative value just by choosing an upper limit of the desired energy spectrum, say $E_{\max}$. Now, since *N* is the largest integer less than $-\mu-\frac{1}{2}$, then using (9) we obtain

$$N = \left\lfloor (E_{\max}/2\omega)-\tfrac{1}{2}\right\rfloor, \quad (10)$$

where $\lfloor x \rfloor$ stands for the largest integer less than *x*. In the following section, we present the TRA solution of the problem, which is written in terms of a new polynomial defined in Ref. [19] by its three-term recursion relation and initial values.

## 3. TRA solution of the problem

With the basis parameters $\mu$ given by (9), the action of the wave operator on the basis elements (8) reduces to the following

$$\mathcal{D}\phi_n(x) = -\frac{\omega}{2}x^{\mu+\frac{5}{4}}e^{-1/2x}\left[(2n+2\mu+1)^2-\left(\ell+\tfrac{1}{2}\right)^2-(\omega a^2)x\right]Y_n^{\mu}(x). \quad (11)$$

Substituting this action in the wave equation $\mathcal{D}\psi(r)=\sum_n f_n \mathcal{D}\phi_n(x)=0$ and using the recursion relation of the Bessel polynomial (A2), we obtain the following three-term recursion relation for the expansion coefficients

$$\begin{aligned}\left(\ell+\tfrac{1}{2}\right)^2 F_n &= \left[\left(2n+2\mu+1\right)^2+\frac{\mu\omega a^2/2}{(n+\mu)(n+\mu+1)}\right]F_n \\ &+\frac{\omega a^2}{2}\left[\frac{n+1}{(n+\mu+1)(2n+2\mu+3)}F_{n+1}-\frac{n+2\mu}{(n+\mu)(2n+2\mu-1)}F_{n-1}\right]\end{aligned} \tag{12}$$

where we have written $f_n = f_0 F_n$ making $F_0=1$. If we define $P_n=\frac{(-1)^n n!\left(2\mu+1\right)}{\left(2n+2\mu+1\right)\left(2\mu+1\right)_n}F_n$, where the Pochhammer symbol (a.k.a. shifted factorial) $\left(a\right)_n = a(a+1)(a+2)...(a+n-1)=\frac{\Gamma(n+a)}{\Gamma(a)}$, then this recursion becomes

$$\begin{aligned}\left(\ell+\tfrac{1}{2}\right)^2 P_n &= \left[\left(2n+2\mu+1\right)^2+\frac{\mu\omega a^2/2}{(n+\mu)(n+\mu+1)}\right]P_n \\ &+\frac{\omega a^2}{2}\left[\frac{n}{(n+\mu)\left(2n+2\mu+1\right)}P_{n-1}-\frac{n+2\mu+1}{(n+\mu+1)\left(2n+2\mu+1\right)}P_{n+1}\right]\end{aligned} \tag{13}$$

Comparing this recursion relation to that of the polynomial $B_n^\mu(z;\gamma)$ defined in [19] by its three-term recursion relation, which is shown here in the Appendix as (A10), we conclude that $P_n = B_n^\mu(z;\gamma)$ where

$$\gamma=-16/\omega a^2, \qquad z=-\left(4/\omega a^2\right)\left(\ell+\tfrac{1}{2}\right)^2. \tag{14}$$

Finally, the kth bound state wavefunction for this spiked oscillator reads

$$\psi_k(r)= f_0(E_k)\left(\omega r^2\right)^{-\mu-\frac{1}{4}} e^{-\omega r^2/2}\sum_{n=0}^{N} G_n B_n^\mu(z;\gamma) Y_n^\mu(1/\omega r^2), \tag{15}$$

where $G_n=\left(2n+2\mu+1\right)\left(2\mu+1\right)_n/(-1)^n n!\left(2\mu+1\right)$ and $N=\left\lfloor\left(E_k/2\omega\right)-\tfrac{1}{2}\right\rfloor$. For a given set of physical parameters $\left\{\omega,a,\ell\right\}$ and bound state energy $E_k$, the basis parameter $\mu$ is given by (9) whereas $z$ and $\gamma$ are given by (14). All physical properties of the system (e.g., energy spectrum of the bound states, density of states, etc.) are obtained from the properties of the polynomial $B_n^\mu(z;\gamma)$ (e.g., its weight function, generating function, asymptotics, zeros, etc.). For example, the energy spectrum could easily be obtained from the spectrum formula of this polynomial. Unfortunately, the analytic properties of $B_n^\mu(z;\gamma)$ are not yet known. It remains an open problem in orthogonal polynomials along with other similar problems. For an expose of such open problems, one could study section 5 of [20] and references therein. Therefore, we are forced to resort to numerical means to calculate the energy spectrum of the system. In the following section, we use the "potential parameter spectrum" (PPS) technique [21] to calculate the bound states energy eigenvalues $\left\{E_k\right\}$ for a given set of physical parameters $\left\{\omega,a,\ell\right\}$.

## 4. Energy spectrum

Since the analytic properties of the TRA polynomial $B_n^{\mu}(z;\gamma)$ in the wavefunction expansion (15) is not yet known, we obtain a numerical solution of its associated three-term recursion relation (13). The aim is to find as many of the binding energies in this spiked oscillator potential (4) as desired and with high enough accuracy. We do that by writing the *symmetric* version of the recursion relation (13) as an eigenvalue equation $y|P\rangle = T|P\rangle$ where *T* is the $(N+1)\times(N+1)$ symmetric tridiagonal matrix $T_{n,m} = a_n\delta_{n,m} + b_{n-1}\delta_{n,m+1} + b_n\delta_{n,m-1}$ with

$$a_n = \left(2n+2\mu+1\right)^2 + \frac{\mu\omega a^2/2}{(n+\mu)(n+\mu+1)}, \tag{16a}$$

$$b_n = \frac{\omega a^2/2}{n+\mu+1}\sqrt{\frac{-(n+1)(n+2\mu+1)}{(2n+2\mu+1)(2n+2\mu+3)}}. \tag{16b}$$

The problem in finding the energy spectrum from the eigenvalue equation $y|P\rangle = T|P\rangle$ is that the matrix elements of *T* do depend on the energy itself via the parameter $\mu$ as shown in (9). However, out of all energies that enter in the construction of *T*, the energy spectrum are only those that produce the eigenvalue $y = \left(\ell+\frac{1}{2}\right)^2$ for a given angular momentum $\ell$. The technique that utilizes this idea to produce the energy spectrum is the PPS [21] and it goes as follows:

1. For a given set of values of the physical parameters $\{\omega, a\}$, we choose an energy interval $0 \le E \le E_{\max}$ with the desired $E_{\max}$ and select *M* points in the interval $\{E_i\}_{i=1}^{M}$.
2. For the energy $E_1$, we calculate the recursion coefficients $\{a_n, b_n\}_{n=0}^{N}$ using (16) and (9) with *N* given by (10). Then, we compute the eigenvalues of the tridiagonal matrix *T* in $y|P\rangle = T|P\rangle$ as the set $\{y_n(E_1)\}_{n=0}^{N}$ sorted in ascending order.
3. We repeat step 2 above for $E_2$ to get the sorted set $\{y_n(E_2)\}_{n=0}^{N}$ and continue this for all energy points in the interval up to $E_M$ that gives the sorted set $\{y_n(E_M)\}_{n=0}^{N}$.
4. For $n=0$, we make a functional fit of the *M* points $\{E_i\}_{i=1}^{M}$ versus $\{y_0(E_i)\}_{i=1}^{M}$ and call this function $E_0(y)$. The ground state energy will then be $E_0\left(y = \left(\ell+\frac{1}{2}\right)^2\right)$.
5. We repeat step 4 for $n=1$ where we fit the *M* points $\{E_i\}_{i=1}^{M}$ versus $\{y_1(E_i)\}_{i=1}^{M}$ to get the function $E_1(y)$ giving the first excited state energy as $E_1\left(y = \left(\ell+\frac{1}{2}\right)^2\right)$. We continue till $n=N$ that gives $E_N\left(y = \left(\ell+\frac{1}{2}\right)^2\right)$.

For functional fitting, we used the Haymaker-Schlessinger continued fraction routine with fitting order *M* [22]. For a given choice of the physical parameters $\{\omega, a\}$ and for several angular momenta, we use the above PPS procedure to calculate the lowest bound states energies. In Table 1, we give the deviation of these energies from that of the pure isotropic oscillator. That is, in Table 1, we list the energy deviations

$$\Delta E_n := E_n - \omega\left(2n+\ell+\tfrac{3}{2}\right). \tag{17}$$

Figure 1 is a plot of the wavefunction (15) for the lowest bound states corresponding to the $\ell = 5$ column of Table 1.

## 5. Independent verification of results

It has been demonstrated elsewhere that the PPS technique *may not* produce accurate enough results in a finite basis like (7) [23]. Therefore, in this section we try to obtain an independent numerical evaluation of the energy spectrum using two approximation methods.

### 5.1 The first method:

This is an approximate analytical calculation of the roots of the determinant of the $(N+1)\times(N+1)$ tridiagonal symmetric matrix $T-\left(\ell+\frac{1}{2}\right)^2 I$ for a given set of physical parameters $\{\omega, a, \ell\}$, where *I* is the $(N+1)\times(N+1)$ unit matrix. If we call this determinant $D_{N+1}(\mu)$ then it is well-known that such a determinant could be evaluated recursively as follows

$$D_{N+1} = \begin{vmatrix} A_0 & B_0 & & & & & \\ B_0 & A_1 & B_1 & & & & \\ & B_1 & A_2 & B_2 & & & \\ & & .. & .. & .. & & \\ & & & .. & .. & .. & \\ & & & & B_{N-2} & A_{N-1} & B_{N-1} \\ & & & & & B_{N-1} & A_N \end{vmatrix} = A_N D_N - B_{N-1}^2 D_{N-1}, \tag{18}$$

where $A_n(\mu) = a_n - \left(\ell+\frac{1}{2}\right)^2$ and $B_n(\mu) = b_n$. The initial values for the solution of the recursion (18) is $D_1 = A_0$ and $D_2 = A_0 A_1 - B_0^2$. We computed the roots $\{\mu_n\}_{n=0}^N$ of the determinant $D_{N+1}(\mu)$ for several values of matrix size *N*. In Table 2, we list $\{\Delta E_n\}_{n=0}^N$ defined by (17) for several values of *N* where $\{E_n = -2\omega\mu_n\}_{n=0}^N$. The Table shows excellent convergence to the values in Table 1 for the chosen value of $\ell$. For evaluating the roots of $D_{N+1}(\mu)$, we used the built-in **root** function in the programming software Mathcad®. We obtained similar superb match with Table 1 for all other values of $\ell$.

### 5.2 The second method:

This method is based on the calculation of the Hamiltonian matrix in a proper and complete basis set whose square-integrable elements are

$$\chi_n(x) = C_n\, x^{\beta} e^{-x/2} L_n^{\nu}(x), \tag{19}$$

where $L_n^{\nu}(x)$ is the Laguerre polynomial with $\nu > -1$, $x = (\lambda r)^2$ and $\lambda$ is a positive scale parameter of inverse length dimension. We choose $C_n = \sqrt{\frac{2\Gamma(n+1)}{\Gamma(n+\nu+1)}}$ and $2\beta = \nu + \frac{1}{2}$ making this basis set orthonormal (i.e., $\langle \chi_n | \chi_m \rangle = \lambda \int_0^\infty \chi_n(x)\chi_m(x) dr = \delta_{n,m}$). To calculate the matrix elements of the Hamiltonian in this basis, $\langle \chi_m | H | \chi_n \rangle$, we need to evaluate the action of the Hamiltonian operator on the basis elements, $H|\chi_n\rangle$, then project on the left by $\langle\chi_m|$ and integrate. In terms of the dimensionless variable $x$, this action reads as follows

$$H|\chi_n\rangle = -\lambda^2 C_n x^\beta e^{-x/2}\left\{-\left(2n+\nu+1\right)+\frac{x}{2}\left[1-\frac{\omega^2}{\lambda^4}\right]+\frac{\nu^2-\left(\ell+\frac{1}{2}\right)^2}{2x}-\frac{\left(\lambda a\right)^2}{2x^2}\right\}L_n^\nu(x), \quad (20)$$

where we have used the differential equation of the Laguerre polynomials. To simplify this action, we choose $\nu=\ell+\frac{1}{2}$. Using the recursion relation and orthogonality of the Laguerre polynomial, we obtain the following matrix elements of the Hamiltonian

$$\langle\chi_n|H|\chi_m\rangle = \frac{\lambda^4 a^2}{2}C_n C_m\int_0^\infty x^{\ell-\frac{3}{2}}e^{-x}L_n^{\ell+\frac{1}{2}}(x)L_m^{\ell+\frac{1}{2}}(x)dx+\frac{\lambda^2}{2}\left(1+\frac{\omega^2}{\lambda^4}\right)\left(2n+\ell+\tfrac{3}{2}\right)\delta_{n,m}$$
$$\frac{\lambda^2}{2}\left(1-\frac{\omega^2}{\lambda^4}\right)\left[\delta_{n,m+1}\sqrt{n\left(n+\ell+\tfrac{1}{2}\right)}+\delta_{n,m-1}\sqrt{\left(n+1\right)\left(n+\ell+\tfrac{3}{2}\right)}\right] \quad (21)$$

To simplify even further, we could choose $\lambda^2=\omega$ in which case we obtain the following elements of the Hamiltonian matrix

$$\langle\chi_n|H|\chi_m\rangle = \omega\left(2n+\ell+\tfrac{3}{2}\right)\delta_{n,m}+\frac{\left(\omega a\right)^2}{2}C_n C_m\int_0^\infty x^{\ell-\frac{3}{2}}e^{-x}L_n^{\ell+\frac{1}{2}}(x)L_m^{\ell+\frac{1}{2}}(x)dx. \quad (22)$$

To get a highly accurate evaluation of the integral in (21) or (22), we used Gauss quadrature approximation associated with the Laguerre polynomials (see, for example, Appendix B in [12] or Appendix A in [25]). Rigorously, integrability dictates that the exponent $\ell-\frac{3}{2}$ of $x$ inside the integral be greater than $-1$. Thus, we expect computational difficulty only for $\ell=0$. However, numerically we can improve accuracy of the result of integration by increasing the size of the basis (19) and/or choosing a proper value of the scale parameter $\lambda$ (for example, choosing $\lambda^2=\omega$). We calculated the bound states energies $\{E_k\}_{k=0}^{M-1}$ as the eigenvalues of the $M\times M$ Hamiltonian matrix (21) for the same physical parameters as those in Table 1. In Table 3, we list $\{\Delta E_n\}$ defined by (17) for the lowest bound states. The good agreement between Table 1 and Table 3 improves with an increasing angular momentum since then the system is pushed away (by the increased orbital motion) from the origin making the $1/r^4$ singularity less detrimental to the calculation.

Finally, in Table 4 we compare our numerical results for the energy spectrum to those obtained by A. K. Roy [12].

# 6. Conclusion

The model presented in this work by the isotonic oscillator potential (4) belongs to the singular versions of (1) with the highest $1/r^4$ singularity. In the literature, this potential is also known as the "spiked harmonic oscillator". The solution of this potential model, which is a non-trivial task, is exceedingly relevant to various physical scenarios. Most of the known methods of solution of the Schrödinger equation did not succeed in yielding an exact solution for this problem. Nonetheless, we were able to obtain the exact solution given by Eq. (15) using the TRA. The most important advantage of this solution is that it is a finite series with relatively small number of terms. In fact, the number of terms for the lowest six bound states shown in Fig. 1 are 3, 4,…,7 (i.e., $N=k+2$). This feature overcomes the difficulty in finding the complete analytic properties of the new TRA polynomial $B_n^\mu(z;\gamma)$ because we can easily write them explicitly for these low degrees just by using the recursion relation (A10).

## Acknowledgments:

I am grateful to I. Assi for comments and fruitful discussions. The help provided by S. Al-Marzoug in literature search is highly appreciated.

## Appendix: Bessel polynomial on the real line

The Bessel polynomial on the positive real line is defined in terms of the hypergeometric functions as follows (see section 9.13 of the book by Koekoek *et. al* [26] but make the replacement $x\mapsto 2x$ and $a\mapsto 2\mu$)

$$Y_n^\mu(x)={}_2F_0\left(\left.{-n,n+2\mu+1 \atop —}\right|-x\right)=\left(n+2\mu+1\right)_n x^n\,{}_1F_1\left(\left.{-n \atop -2(n+\mu)}\right|1/x\right), \tag{A1}$$

where $x\geq 0$, $n=0,1,2,..,N$ and $N$ is a non-negative integer. The real parameter $\mu$ is negative such that $\mu<-N-\frac{1}{2}$. The Pochhammer symbol $(a)_n$ (a.k.a. shifted factorial) is defined as $\left(a\right)_n=a(a+1)(a+2)...(a+n-1)=\frac{\Gamma(n+a)}{\Gamma(a)}$. The Bessel polynomial could also be written in terms of the Laguerre polynomial as: $Y_n^\mu(x)=n!\left(-x\right)^n L_n^{-(2n+2\mu+1)}\left(1/x\right)$. The three-term recursion relation reads as follows:

$$\begin{aligned}2xY_n^\mu(x)&=\frac{-\mu}{(n+\mu)(n+\mu+1)}Y_n^\mu(x)\\&-\frac{n}{(n+\mu)(2n+2\mu+1)}Y_{n-1}^\mu(x)+\frac{n+2\mu+1}{(n+\mu+1)(2n+2\mu+1)}Y_{n+1}^\mu(x)\end{aligned} \tag{A2}$$

Note that the constraints on $\mu$ and on the maximum polynomial degree make this recursion definite (i.e., the signs of the two recursion coefficients multiplying $Y_{n\pm1}^\mu(x)$ are the same). Otherwise, these polynomials will not be orthogonal on the real line but on the unit circle in the complex plane. The orthogonality relation reads as follows

$$\int_0^\infty x^{2\mu}e^{-1/x}Y_n^\mu(x)Y_m^\mu(x)\,dx=-\frac{n!\Gamma(-n-2\mu)}{2n+2\mu+1}\delta_{nm}\,. \tag{A3}$$

The differential equation is

$$\left\{x^2\frac{d^2}{dx^2}+\left[1+2x\left(\mu+1\right)\right]\frac{d}{dx}-n\left(n+2\mu+1\right)\right\}Y_n^\mu(x)=0\,. \tag{A4}$$

The forward and backward shift differential relations read as follows

$$\frac{d}{dx}Y_n^\mu(x)=n\left(n+2\mu+1\right)Y_{n-1}^{\mu+1}(x). \tag{A5}$$

$$x^2\frac{d}{dx}Y_n^\mu(x)=-\left(2\mu x+1\right)Y_n^\mu(x)+Y_{n+1}^{\mu-1}(x). \tag{A6}$$

We can write $Y_{n+1}^{\mu-1}(x)$ in terms of $Y_n^\mu(x)$ and $Y_{n\pm1}^\mu(x)$ as follows

$$\begin{gathered}2Y_{n+1}^{\mu-1}(x)=\frac{(n+1)(n+2\mu)}{(n+\mu)(n+\mu+1)}Y_n^\mu(x)\\ +\frac{n(n+1)}{(n+\mu)(2n+2\mu+1)}Y_{n-1}^\mu(x)+\frac{(n+2\mu)(n+2\mu+1)}{(n+\mu+1)(2n+2\mu+1)}Y_{n+1}^\mu(x)\end{gathered} \tag{A7}$$

Using this identity and the recursion relation (A2), we can rewrite the backward shift differential relation (A6) as follows

$$\begin{gathered}2x^2\frac{d}{dx}Y_n^\mu(x)=n(n+2\mu+1)\times\\ \left[-\frac{Y_n^\mu(x)}{(n+\mu)(n+\mu+1)}+\frac{Y_{n-1}^\mu(x)}{(n+\mu)(2n+2\mu+1)}+\frac{Y_{n+1}^\mu(x)}{(n+\mu+1)(2n+2\mu+1)}\right]\end{gathered} \tag{A8}$$

The generating function is

$$\sum_{n=0}^\infty Y_n^\mu(x)\frac{t^n}{n!}=\frac{2^{2\mu}}{\sqrt{1-4xt}}\left(1+\sqrt{1-4xt}\right)^{-2\mu}\exp\left[2t/(1+\sqrt{1-4xt})\right]. \tag{A9}$$

The polynomial $B_n^\mu(z;\gamma)$ is defined in Ref. [19] by its three-term recursion relation, Eq. (15) therein, which reads

$$\begin{gathered}z\,B_n^\mu(z;\gamma)=\left[\frac{-2\mu}{(n+\mu)(n+\mu+1)}+\gamma\left(n+\mu+\tfrac{1}{2}\right)^2\right]B_n^\mu(z;\gamma)\\ -\frac{n}{(n+\mu)\left(n+\mu+\frac{1}{2}\right)}B_{n-1}^\mu(z;\gamma)+\frac{n+2\mu+1}{(n+\mu+1)\left(n+\mu+\frac{1}{2}\right)}B_{n+1}^\mu(z;\gamma)\end{gathered} \tag{A10}$$

where $B_0^\mu(z;\gamma)=1$ and $B_{-1}^\mu(z;\gamma):=0$.

## Tables Caption:

**Table 1**: PPS calculation (in atomic units) of the energy deviations from the pure isotropic oscillator $\{\Delta E_n\}_{n=0}^{N}$ as defined by (17) for the lowest bound states energies and for several values of the angular momentum. We took $\omega = 1.0$ and $a = 0.5$ . The number of fitting energy points is $M = 100$ .

**Table 2**: The energy deviations (in atomic units) $\{\Delta E_n\}_{n=0}^{N}$ defined by (17) for the lowest bound states calculated using the roots of the determinant (18) for several matrix sizes $(N+1)$ . We took $\omega = 1.0$ , $a = 0.5$ and $\ell = 5$ .

**Table 3**: The energy deviations (in atomic units) $\{\Delta E_n\}$ defined by (17) for the lowest bound states where the energy spectrum $\{E_n\}$ is obtained as the eigenvalues of the Hamiltonian matrix (21). We took the same physical parameters as in Table 1 and chose $\lambda^2 = \omega$. The Hamiltonian matrix size $M = 100$ .

**Table 4**: The lowest three bound state energies (in atomic units) obtained using the TRA compared to those obtained by A. K. Roy in Table 6 of [12] using the "imaginary-time evolution method". The oscillator frequency $\omega = 1.0$ .

## Figure Caption:

**Fig. 1**: Plots of the wavefunction given by the finite series (15) for the lowest six bound states corresponding to the $\ell = 5$ column of Table 1. The values of the physical parameters are the same as in Table 1. The wavefunctions are un-normalized and the horizontal axis is the radial coordinate *r* (in units of $1/\sqrt{\omega}$ ).

**Table 1**

| $n$ | $\ell = 3$ | $\ell = 4$ | $\ell = 5$ | $\ell = 6$ | $\ell = 7$ |
|---|---|---|---|---|---|
| 0 | 0.014093664 | 0.007895881 | 0.005038139 | 0.003491750 | 0.002561967 |
| 1 | 0.020199526 | 0.010734904 | 0.006579901 | 0.004420052 | 0.003163657 |
| 2 | 0.026201793 | 0.013558855 | 0.008118140 | 0.005347298 | 0.003764967 |
| 3 | 0.032153953 | 0.016363232 | 0.009652883 | 0.006273491 | 0.004365899 |
| 4 | 0.038047366 | 0.019160964 | 0.011184153 | 0.007198635 | 0.004966453 |
| 5 | 0.043892069 | 0.021940815 | 0.012711974 | 0.008122733 | 0.005566631 |
| 6 | 0.049552918 | 0.024707753 | 0.014236371 | 0.009045789 | 0.006166432 |
| 7 | 0.055291222 | 0.027459101 | 0.015757368 | 0.009967806 | 0.006765857 |
| 8 | 0.060543550 | 0.030197819 | 0.017274988 | 0.010888788 | 0.007364908 |
| 9 | 0.065805452 | 0.032920750 | 0.018789252 | 0.011808738 | 0.007963584 |

**Table 2**

| $n$ | $N = 0$ | $N = 1$ | $N = 2$ | $N = 5$ | $N = 10$ |
|---|---|---|---|---|---|
| 0 | 0.005042540 | 0.005038139 | 0.005038139 | 0.005038139 | 0.005038139 |
| 1 | | 0.006590264 | 0.006579901 | 0.006579901 | 0.006579901 |
| 2 | | | 0.008136005 | 0.008118140 | 0.008118140 |
| 3 | | | | 0.009652883 | 0.009652883 |
| 4 | | | | 0.011184159 | 0.011184152 |
| 5 | | | | 0.012761400 | 0.012711974 |
| 6 | | | | | 0.014236371 |
| 7 | | | | | 0.015757367 |
| 8 | | | | | 0.017274987 |
| 9 | | | | | 0.018789252 |

**Table 3**

| $n$ | $\ell = 3$ | $\ell = 4$ | $\ell = 5$ | $\ell = 6$ | $\ell = 7$ |
|---|---|---|---|---|---|
| 0 | 0.014088829 | 0.007895857 | 0.005038137 | 0.003491750 | 0.002561967 |
| 1 | 0.020168585 | 0.010734771 | 0.006579892 | 0.004420052 | 0.003163657 |
| 2 | 0.026152741 | 0.013558349 | 0.008118108 | 0.005347296 | 0.003764967 |
| 3 | 0.032044738 | 0.016366668 | 0.009652793 | 0.006273485 | 0.004365898 |
| 4 | 0.037847372 | 0.019159733 | 0.011183941 | 0.007198618 | 0.004966452 |
| 5 | 0.043562902 | 0.021937470 | 0.012711533 | 0.008122694 | 0.005566627 |
| 6 | 0.049193129 | 0.024699726 | 0.014235533 | 0.009045709 | 0.006166424 |
| 7 | 0.054739450 | 0.027446263 | 0.015755884 | 0.009967653 | 0.006765841 |
| 8 | 0.060202910 | 0.030176757 | 0.017272505 | 0.010888512 | 0.007364876 |
| 9 | 0.065584229 | 0.032890796 | 0.018785288 | 0.011808267 | 0.007963526 |

**Table 4**

| | $a^2 = 0.001$ | | $a^2 = 1.0$ | |
|---|---|---|---|---|
| $\ell$ | This work | Ref. [12] | This work | Ref. [12] |
| 3 | 4.50005714 | 4.50005713 | 4.55432941 | 4.55432930 |
| | 6.50008253 | 6.50008253 | 6.57618636 | 6.57618592 |
| | 8.50010792 | 8.50010792 | 8.59705146 | 8.59705033 |
| 4 | 5.50003175 | 5.50003174 | 5.53112085 | 5.53112085 |
| | 7.50004329 | 7.50004328 | 7.54196630 | 7.54196634 |
| | 9.50005483 | 9.50005483 | 9.55260775 | 9.55260789 |
| 5 | 6.50002020 | 6.50002020 | 6.52000759 | 6.52000759 |
| | 8.50002642 | 8.50002641 | 8.52603424 | 8.52603424 |
| | 10.50003263 | 10.5000326 | 10.53200890 | 10.5320089 |
| 10 | 11.50000501 | 11.5000050 | 11.50500707 | 11.5050070 |
| | 13.50000588 | 13.5000058 | 13.50587573 | 13.5058757 |
| | 15.50000676 | 15.5000067 | 15.50674384 | 15.5067438 |
| 40 | 41.50000031 | 41.5000003 | 41.50031254 | 41.5003125 |
| | 43.50000033 | 43.5000003 | 43.50032761 | 43.5003276 |
| | 45.50000034 | 45.5000003 | 45.50034267 | 45.5003426 |

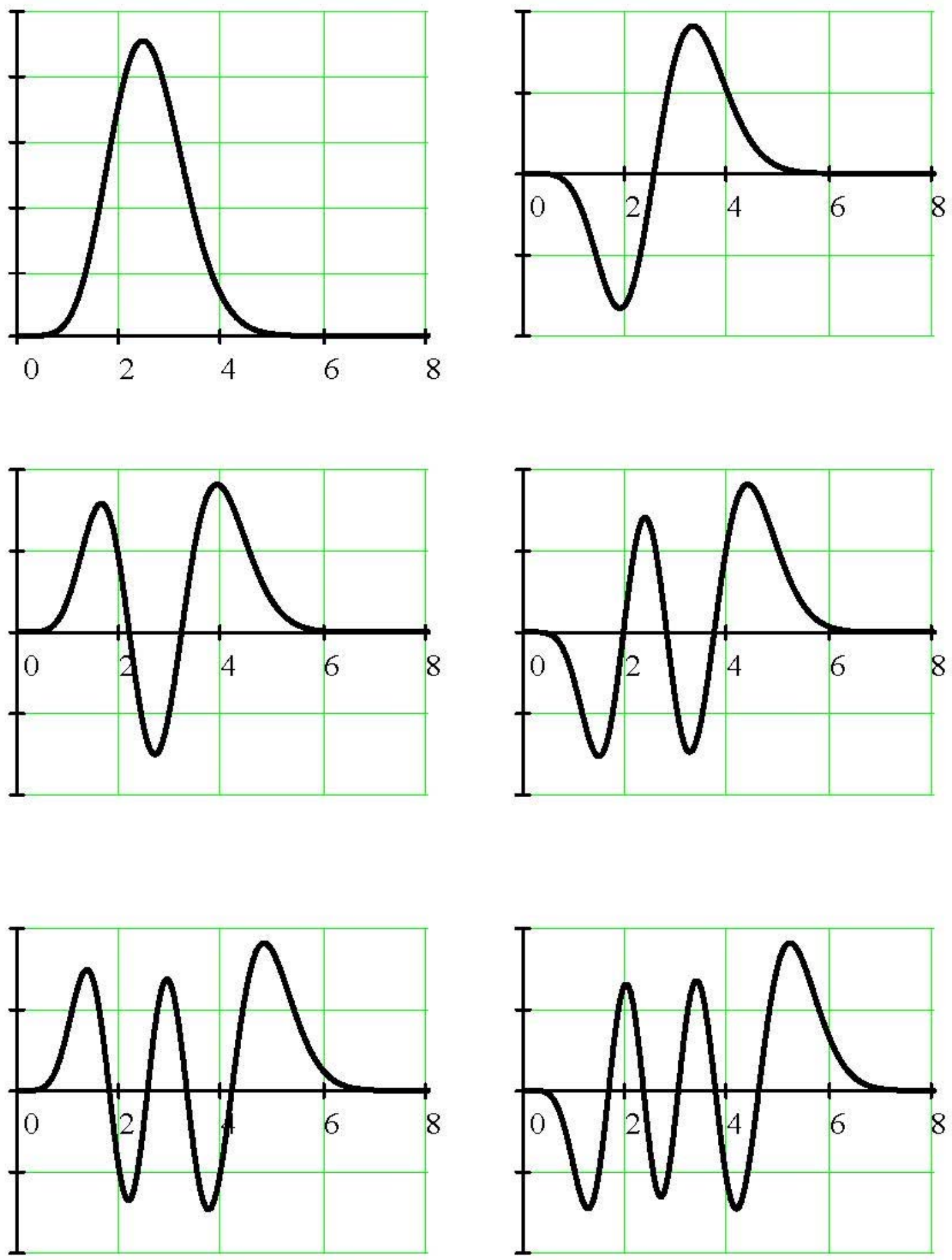

**Fig. 1**